\documentstyle[12pt]{article}

\begin{document}

\title{How to account for virtual arbitrage in the standard derivative
pricing} 
\author{ Kirill Ilinski \thanks{E-mail:  kni@th.ph.bham.ac.uk} \\
[0.3cm] 
{\small\it IPhys Group, CAPE,
14-th line of Vasilievskii's Island, 29}\\
{\small\it St-Petersburg, 199178, Russian Federation}
\\ [0.1cm]
{\small and} \\ [0.1cm]
{\small\it School of Physics and Space Research,
University of Birmingham,} \\
{\small\it Edgbaston B15 2TT, Birmingham, United Kingdom}
}

\vspace{3cm}

\date{ }

\vspace{1cm}

\maketitle

\begin{abstract} In this short note we show how virtual arbitrage
opportunities can be modelled and included in the standard derivative
pricing without changing the general framework.  
\end{abstract}

Whatever people say about the drawbacks of the Black-Scholes (BS) approach~\cite{BS}
to derivative pricing, it is a standard method and almost any pricing and
hedging software in financial institutions is based on it.  Practitioners
have got used to BS-like partial differential equations, martingales and
other related mathematical animals.  Both analytical and numerical methods
are well developed and it is hardly surprising that practitioners are
rather reluctant to ``buy" complicated new theories.  That is why it is
interesting to see how some limitations of BS analysis can be overcome in
the same mathematical framework without disturbing the foundations.

One way to improve BS is to use a more realistic price process instead of
the geometrical random walk.  The most popular alternatives are ARCH-GARCH
models where the volatility of the return is assumed to be stochastic.
Although they are a better approximation for the price process 
the description is far from perfect~\cite{Mantegna}. Another line of 
attack is the no-arbitrage
constraint. Some empirical studies have demonstrated existence of the short 
lived arbitrage opportunities~\cite{der,Sofianos}. In this paper we show how 
to generalize BS analysis to the case where the virtual 
arbitrage opportunities exist.

In Ref~\cite{IS} Ilinski and Stepanenko proposed BS-based model to
account for virtual arbitrage. The model substitutes the constant interest
rate $r_0$ as a rate of return on a riskless portfolio by some stochastic 
process $r_0+x(t)$. In the limit of fast enough market reaction to the 
arbitrage opportunities (see~\cite{IS} 
for details) the model is reduced to the Ornstein-Uhlenbeck process 
for the variable $x(t)$:
\begin{equation}
\frac{d x}{dt} + \lambda x = \eta (t)
\label{OU}
\end{equation}
where $\eta (t)$ is the white noise stochastic process
$$
\langle \eta(t) \rangle = 0 \ , \qquad
\langle \eta(t) \eta(t^{\prime}) \rangle = 
\Sigma^2 \delta (t-t^{\prime}) \ .
$$
The parameter $\lambda$ characterises how fast the market reacts to an 
arbitrage opportunity while the parameter $\Sigma^2$ defines how often such
opportunities appear and how profitable they are. In the limit 
$\Sigma^2\rightarrow 0$ or $\lambda\rightarrow \infty$ we recover the results 
of the BS analysis.

As was already mentioned, real price processes are not geometrical random
walks. However, if we assume they are, the parameters $\Sigma^2$ and
$\lambda$ can be estimated from the market data as
$$
\frac{\Sigma^2}{2\lambda} = 
\left\langle\left(
\frac{{\cal L}_{\rm BS} V}{V - S\frac{\partial V}{\partial S}}
\right)^{2}(t)\right\rangle_{\rm market}
$$
and
$$
\lambda = -\frac{1}{(t - t^{\prime})}\log \left[
\left\langle\frac{{\cal L}_{\rm BS} V}{V - S\frac{\partial V}{\partial S}}(t)
\frac{{\cal L}_{\rm BS} V}{V - S\frac{\partial V}{\partial S}}(t^{\prime})
\right\rangle_{\rm market} \left/  
\left\langle\left(
\frac{{\cal L}_{\rm BS} V}{V - S\frac{\partial V}{\partial S}}
\right)^{2}(t)\right\rangle_{\rm market}\right.
\right] \ ,
$$
with the Black-Scholes operator ${\cal L}_{\rm BS}$ defined as usual:
$$
{\cal L}_{\rm BS} = \frac{\partial }{\partial t} + 
\frac{\sigma^2 S^2}{2}\frac{\partial ^2 }{\partial S^2}  + r_0
S\frac{\partial }{\partial S} - r_0 \ .
$$
Here $S$ is an underlying asset price and $\sigma$ is its volatility.
The (historic) market values of the parameters allow us to estimate how
appropriate the no-arbitrage approximation is, {\it i.e.} when it 
is possible to neglect the correction terms come from the virtual arbitrage.
To first order with respect to the parameter $\frac{\Sigma^2}{2\lambda}$
the correction to the BS equation was found in Ref~\cite{IS} together with 
explicit formulas for the call and put vanilla options. However it is clear 
that in the case $\frac{\Sigma^2}{2\lambda} \geq 1$ this expansion does 
not make much sense and we have to find a way to sum over all relevant terms.

How real is this possibility?
Let us assume that the arbitrage
on the riskless portfolio $V- S\frac{\partial V}{\partial S}$ appears every day
and is on average about one per cent (in absolute value). We assume also that 
the arbitrage is washed out during the day and there is no interference between
different arbitrage opportunities. Then the parameter $\lambda$ is about 260,
the arbitrage produces the return of about 260 per cent per year 
and the parameter $\frac{\Sigma^2}{2\lambda}$ is about $6.76=(2.6)^2>>1$. 
In this case the no-arbitrage BS pricing formulas are too 
crude an approximation. If however the 
arbitrage appears only once a week then $\frac{\Sigma^2}{2\lambda}\sim 0.25$ 
and the parameter can be used as an expansion parameter. These examples
show that accounting for the virtual arbitrage can sometimes be very
appropriate. It is worth mentioning that here we have neglected the bid-ask spread 
and transaction costs for illustrative purposes. In reality both effects 
have to be included.

It is not difficult to derive an analogue of the BS equation for the derivative
price $\bar{U}(S,t)$ averaged over virtual arbitrage process (\ref{OU}):
\begin{equation}
\bar{U}(S,t) = \int dx \bar{u}(x,S,t)
\label{u}
\end{equation}
where the function $\bar{u}(x,S,t)$ is the solution of the problem:
\begin{equation}
\frac{\partial \bar{u}}{\partial t} + 
\frac{\sigma^2 S^2}{2}\frac{\partial ^2 \bar{u}}{\partial S^2}  + (r_0 +x)
S\frac{\partial \bar{u}}{\partial S} - (r_0+x)\bar{u} +
\frac{\Sigma^2}{2}\frac{\partial ^2 \bar{u}}{\partial x^2} +
\lambda \frac{\partial x \bar{u}}{\partial x} = 0 \ ,
\label{ur1}
\end{equation}
$$
\bar{u}(x,S,t)|_{t=T} = \delta (x) \cdot Payoff (S) \ .
$$
A concise informal derivation of the equation is in the Appendix.
It is straightforward to check that $\bar{U}(S,t)$ converges to the 
solution of the BS equation as $\Sigma^2$ or $1/\lambda$ tend to zero.
It has to be used for both pricing and hedging, allows  
standard numerical methods to be used in its calculation 
and can serve as a basis for further generalizations.

Two such generalizations are immediately obvious:
\begin{enumerate}
\item It is possible to use another form of the stabilizing market reaction, if
it fits the observable data better. In this case the process will be defined as
\begin{equation}
\frac{d x}{dt} + f(x) = \eta (t)
\label{NOU}
\end{equation}
with the same white noise $\eta (t)$
$$
\langle \eta (t)\rangle = 0 \ , \qquad
\langle \eta(t) \eta(t^{\prime}) \rangle = 
\Sigma^2 \delta (t-t^{\prime}) \ ,
$$
which leads to the following problem for the function $\bar{u}(x,S,t)$:
\begin{equation}
\frac{\partial \bar{u}}{\partial t} + 
\frac{\sigma^2 S^2}{2}\frac{\partial ^2 \bar{u}}{\partial S^2}  + (r_0 +x)
S\frac{\partial \bar{u}}{\partial S} - (r_0+x)\bar{u} +
\frac{\Sigma^2}{2}\frac{\partial ^2 \bar{u}}{\partial x^2} +
\frac{\partial f(x) \bar{u}}{\partial x} = 0\ ,
\label{ur2}
\end{equation}
$$
\bar{u}(x,S,t)|_{t=T} = \delta (x) \cdot Payoff (S) \ .
$$
The average price $\bar{U}(S,t)$ then is found as before from Eqn(\ref{u}).

\item In a similar manner it is possible to improve the 
stochastic volatility models~\cite{Merton,HW,Wang}, the jump 
diffusion model~\cite{Merton1} and others. For example, in the case of
the price process $S$ with stochastic volatility $\sigma$:
$$
dS = \phi S dt + \sigma S dw \ , \qquad V\equiv \sigma^2 \ , \quad
dV = \mu V dt + \xi V dz
$$
such that the Wiener processes have correlation $\rho$ and the volatility 
process has zero systematic risk~\cite{HW}, the pricing equation takes 
the form:
$$
\frac{\partial \bar{u}}{\partial t} + 
\frac{1}{2}\left(\sigma^2 S^2\frac{\partial ^2 \bar{u}}{\partial S^2}  + 
2\rho \sigma^3 \xi S \frac{\partial ^2 \bar{u}}{\partial S \partial V} +
\xi^2 V^2\frac{\partial ^2 \bar{u}}{\partial V^2}\right) +
(r_0 +x)
S\frac{\partial \bar{u}}{\partial S} 
+\mu\sigma^2 \frac{\partial \bar{u}}{\partial V} 
$$
$$
- (r_0+x)\bar{u} +
\frac{\Sigma^2}{2}\frac{\partial ^2 \bar{u}}{\partial x^2} +
\lambda \frac{\partial x \bar{u}}{\partial x} = 0 \ ,
$$
with the corresponding final condition:
$$
\bar{u}(x,S,t)|_{t=T} = \delta (x) \cdot Payoff (S) \ .
$$
The average value of the price is given by Eqn(\ref{u}). 
\end{enumerate}

It is possible to combine these generalizations or expand the treatment to 
other BS-like equations. In any case the choice of one variant or another
should involve extensive empirical study. It is important to 
emphasize however that all the pricing ideology, technology of calculations and
numerical methods remain essentially the same so that the 
corresponding software can be easily adapted for the presence of the virtual 
arbitrage.

Lastly we note an apparent analogy between the virtual arbitrage return
and a stochastic interest rate with mean reversion 
(I am grateful to Neil Johnson and Stefan Thurner for this comment). However, 
as it can be seen from the numeric values, the interest rate time-scale is 
much larger while
the characteristic fluctuation size is considerably smaller than the virtual 
arbitrage counterparts. This makes the corrections valuable even for short
living derivatives when the stochasticity of the interest rate can be neglected.
Another difference is more theoretically important. In the models with 
stochastic interest rate the risk can be actually hedged using bonds with
various maturities. In the case of the virtual arbitrage this risk is intrinsic
and cannot be hedged. In other words there is no tradable instrument involving 
the same risk which can be used for pricing and hedging. Even the whole concept
of pricing and hedging using duplication becomes somewhat obscure.

\newpage

\section*{Appendix} In this appendix we give an informal derivation of the
main equation (\ref{ur1}) using the formalism of functional integrals.

We start with the Black-Scholes equation with the rate of return  on the 
riskless portfolio $V-S\frac{\partial V}{\partial S}$ given by $r_0+x$:
$$
\frac{\partial V}{\partial t} + 
\frac{\sigma^2 S^2}{2}\frac{\partial ^2 V}{\partial S^2}  + (r_0 +x)
S\frac{\partial V}{\partial S} - (r_0+x)V = 0 \ ,
$$
$$
V(S,t)|_{t=T} = \delta(S-S^{\prime}) \ .
$$
To simplify the calculation we change variables as $\tau=T-t$ and $y=ln\ S$
which casts the previous equations in the form:
$$
\frac{\partial V}{\partial \tau} = 
\frac{\sigma^2}{2}\frac{\partial ^2 V}{\partial y^2}  + 
(r_0 +x -\frac{\sigma^2}{2})\frac{\partial V}{\partial y} - (r_0+x)V \ ,
$$
$$
V(y,\tau)|_{\tau=0} = \frac{1}{y^{\prime}}\delta(y-y^{\prime}) \ .
$$
The solution of the problem can be expressed as the following functional 
integral~\cite{Popov}:
$$
V(y,\tau) = \frac{1}{y^{\prime}}\int Dq D\frac{p}{2\pi} \ \ e^{
\int_{0}^{\tau} \left(ip\dot{q} -\frac{\sigma^2}{2}p^2 -(r_{0}+x(\tau)) + 
i(r_{0}+x(\tau)-\frac{\sigma^2}{2})p\right) d\tau
}
$$
with the boundary conditions $q(0)=y^{\prime}$, $q(\tau)=y$.
The functional integral form of the solution is extremely convenient
for the purpose of averaging over trajectories $x(\tau)$ since it presents
the dependence in explicit form. We, however, will use another trick.
Let us first of all average the previous expression over realisations of the 
stochastic process $x(\tau)$ with fixed ends, i.e. when $x(\tau)=x$ and
$x(0)=0$ (since there is no arbitrage at the expiration date of the contract
and later). The probabilistic weight of a particular trajectory $x(\tau)$
for the Ornstein-Uhlenbeck process (\ref{OU}) is given by the expression
$$
\int D\frac{\xi}{2\pi} \ \ e^{\int_{0}^{\tau} (i\xi \dot{x} 
-\frac{\Sigma^2}{2}\xi^2 - i \lambda \xi x) d\tau }
$$
which leads to the following result for the conditional average value of the 
contract
$$
\bar{V}(x,y,\tau) = \frac{1}{y^{\prime}}
\int Dq D\frac{p}{2\pi} Dx D\frac{\xi}{2\pi}
\ \ e^{\int_{0}^{\tau} \left(i\xi \dot{x} +
ip\dot{q} -\frac{\sigma^2}{2}p^2 -(r_{0}+x(\tau)) + 
i(r_{0}+x(\tau)-\frac{\sigma^2}{2})p
 -\frac{\Sigma^2}{2}\xi^2 - i \lambda \xi x\right) d\tau }
$$
$$
q(0)=y \ , \quad q(\tau) = y^{\prime} \ , \quad x(0)=0 \ , \quad x(\tau) =x \ .
$$
Now, instead of evaluating these integrals, we find a partial differential 
equation for $\bar{V}(x,y,\tau)$~\cite{Popov}:
$$
\frac{\partial \bar{V}}{\partial t} = 
\frac{\sigma^2}{2}\frac{\partial ^2 \bar{V}}{\partial y^2}  
+ (r_0 +x-\frac{\sigma^2}{2})
\frac{\partial \bar{V}}{\partial y} - (r_0+x)\bar{V} +
\frac{\Sigma^2}{2}\frac{\partial ^2 \bar{V}}{\partial x^2} +
\lambda \frac{\partial x \bar{V}}{\partial x} \ ,
$$
with the initial conditions
$$
\bar{V}(x,y,\tau)|_{\tau=0} = \frac{1}{y^{\prime}} \delta (x) \ .
$$
Returning back to the initial variables $t$ and $S$ we obtain the problem 
(\ref{ur1})
$$
\frac{\partial \bar{V}}{\partial t} + 
\frac{\sigma^2 S^2}{2}\frac{\partial ^2 \bar{V}}{\partial S^2} 
 + (r_0 +x)
S\frac{\partial \bar{V}(x,S,t)}{\partial S} - (r_0+x)\bar{V} +
\frac{\Sigma^2}{2}\frac{\partial ^2 \bar{V}}{\partial x^2} +
\lambda \frac{\partial x \bar{V}}{\partial x} = 0 \ ,
$$
$$
\bar{V}(x,S,t)|_{t=T} = \delta (x) \cdot \delta (S-S^{\prime}) \ .
$$
Integration of the solution over $x$ (to get the unconditional average) and
the convolution with the final payoff complete the consideration.

In a similar way various generalizations of the equations can be derived.
It is clear that the functional integral is a very convenient 
tool for such kind of
manipulations. We would like to note however that the functional ingral 
formalism is full of subtlties which we have not emphasized here. It serves for
fast derivation but not proving the results. The proof can be obtained
in the routine framework of stochastic calculus.


\begin{thebibliography}{99}

\bibitem{BS} F.  Black, M.  Scholes:
``The Pricing of Options and Corporate Liabilities", 
{\em Journal of Political Economy},
{\bf 81} (1973) 637-659;

\bibitem{Mantegna} Though ARCH-GARCH models can fit the
probability distribution function of prices almost perfectly
for a particular time horizon,
they obey different time scaling laws and it destroy the resemblance (See
R.N. Mantegna and H.E. Stanley: {\it Modeling Financial Data: Comparison
of the Truncated Levy Flight and The ARCH(1) and GARCH(1,1) processes}, 
preprint cond-mat/9804126, available at 
http://xxx.lanl.gov/abs/hep-th/9804126);

\bibitem{IS} K.  Ilinski and A.  Stepanenko:  Derivative Pricing with
Virtual Arbitrage, IPHYS working paper IPHYS-98-3;

\bibitem{der} 
F. Black and M. Scholes: ``The Valuation of Option Contracts and a 
Test of Market Efficiency", {\it Journal of Finance},
{\bf 27} (1972) 399-418; D. Galai: ``Tests of Market Efficiency and
the Chicago Board Option Exchange", {\it Journal of Business},
{\bf 50} (1977) 167-197; R.C. Klemkosky and B.G. Resnick:
``Put-Call Parity and Market Efficiency", {\it Journal of Finance},
{\bf 34} (1979) 1141-1155;

\bibitem{Sofianos} G.  Sofianos:  Index Arbitrage Profitability, NYSE
working paper 90-04; {\it J.of Derivatives}, {\bf 1}, N1 (1993);

\bibitem{Merton} R.C.Merton, ``The Theory of Rational Option Pricing",
{\it The Bell Journal of Economics and Management Science}, {\bf 4}, (1973) 
141-183;

\bibitem{HW} J.C.Hull and A.White, ``Pricing of Options on Asset with
Stochastic Volatilities", {\it Journal of Finance}, {\bf 2} (1987) 281-299;

\bibitem{Wang} D.F.Wang: {\it Hedging the risk in the continuous time option 
pricing model with stochastic volatility}, preprint cond-mat/9807066;
available at http://xxx.lanl.gov/abs/cond-mat/9807066;

\bibitem{Merton1} R.C.Merton, ``Option Pricing When Underlying Stock
Returns Are Discontinuous", {\it J. of Financial Economics}, {\bf 3}, (1976) 
124-144;

\bibitem{Popov} V.N. Popov, {\it Functional integrals in quantum 
field theory and statistical physics}, Dordrecht, 1983.

\end{thebibliography}
\end{document}